# NMR study of native defects in PbSe

D. Koumoulis[1], R. E. Taylor[1], D. King Jr.[1] and L-S Bouchard[1,2*]

[1]Department of Chemistry and Biochemistry, University of California, Los Angeles, CA 90095-1569 USA

[2]UCLA California NanoSystems Institute, Los Angeles, CA 90095-1569 USA

Last Revised: August 21, 2014

**Abstract**

While each atom species in PbSe corresponds to a single crystallographic site and transport measurements reveal a single carrier density, $^{207}$Pb NMR reveals a more complicated picture than previously thought comprising three discrete homogeneous carrier components, each associated with *n*- or *p*-type carrier fractions. The origins of these fractions are discussed in terms of electronic heterogeneity of the native semiconductor. The interaction mechanism between nuclear spins and lattice vibrations via fluctuating spin-rotation interaction, applicable to heavy spin-*1/2* nuclei [*Phys. Rev. B* **74**, 214420 (2006)], does not hold. Instead, a higher-order temperature dependence dominates the relaxation pathway. Shallow acceptor states and deep level defects in the midgap explain the complex temperature dependence of the direct band gap.

# I. INTRODUCTION

Lead selenide (PbSe) is a IV-VI narrow-gap semiconductor with a direct band gap ($E_g$) of 0.27-0.31 eV at the $L$ point of the first Brillouin zone [1]. As a self-doped material, its applications have included thermoelectricity, spintronics, ferroelectricity and thermal imaging [2-5]. PbSe has been produced not only in bulk form but also as quantum dots and thin films [6,7]. *n*-type PbSe arises from either interstitial lead or selenium vacancies whereas *p*-type arises from either lead vacancies or interstitial selenium. The recent discovery of topological crystalline insulators [8,9] featuring enhanced transport properties [8-10] has recently thrust PbSe once again at the forefront of condensed matter research. Thus, a thorough understanding of its properties is required.

Defects affect readouts of thermopower, carrier density and resistivity [11] by introducing variability with respect to contact potentials and sample temperature. However, in each case the transport measurements report a single component, which is an average over the entire sample [12]. Average measurements invariably hide subtleties in the electronic structure. For example, multivalley semiconductors (such as PbTe) with isotropic or anisotropic bands possess more than one effective mass. Many native semiconductors cannot be found in defect-free form. Such is the case for PbSe. In this case, averaging hides critical information. For example, high resistivity does not necessarily indicate an undoped material. PbTe also possesses many defects and inhomogeneities, as documented in Refs. [12,13].

Local probes exist which can interrogate electronic properties at the atomic level. NMR spectroscopy reports on electronic properties at the local level through the electron-nuclear hyperfine interaction, making it possible to reveal distinct electronic environments. Consider the case of PbSe, which has a rock salt structure and therefore, only a single resonance would be expected for either NMR-active nuclei. Yet, previous studies have reported two distinct [207]Pb resonances [14,15]. Explications for these resonances are still lacking. In the present study, we report three resonances in the range 170 K to 420 K, revealing additional electronic inhomogeneity in PbSe not previously reported. The electronic properties are spread among three distinct domains associated with three different [207]Pb resonances possessing *n*- and *p*-type character. We found for both nuclei ([77]Se, [207]Pb) a strong temperature dependence of the Knight shift and the spin-lattice relaxation rate ($1/T_1$) for each component with a cusp at 250 K, whereas above 370 K an unexpected resonance arises. The origins of the discrete components are discussed in terms of sample heterogeneity.

## II. EXPERIMENTAL DETAILS

Powder samples of PbSe were obtained from Alfa Aesar (99% purity on metal basis). The powder X-ray diffraction (PXRD) analysis confirms the crystallinity and single-phase nature of the samples (Fig. 1). The Seebeck coefficient measurements were performed on a custom-built instrument with a delta T of approximately 5 K, using type T thermocouples and measuring the voltage through the copper wires of the thermocouples. The error of the measurement is ± 10%. The Seebeck coefficient was measured to be 120 μV/K, indicating a *p*-type semiconductor. As we shall see below, the nature of the carrier type is more complex.

NMR data were acquired with a Bruker DSX-300 spectrometer operating at a frequency of 62.79 MHz for $^{207}$Pb and 57.24 MHz for $^{77}$Se. Static polycrystalline samples of PbSe were placed in a standard Bruker X-nucleus wideline probe with a 5-mm solenoid coil. Each sample was ground with a mortar and pestle to reduce radiofrequency (*RF*) skin-depth effects at these NMR frequencies and used as-is. The $^{207}$Pb *π/2* pulse width was 4.5 μs, and the $^{77}$Se *π/2* pulse width was 4.0 μs. Spectral data were acquired using a spin-echo sequence [*(π/2)$_x$ – τ – (π)$_y$-τ- acquire*] with the echo delay, τ, set to 20 μs. Magic-angle spinning (MAS) spectra were acquired on the same sample with a standard Bruker MAS probe using a 4-mm outside diameter zirconia rotor with a sample spinning rate of 8 kHz. *T$_1$* relaxation times were acquired with a saturation-recovery technique [16]. The $^{207}$Pb and $^{77}$Se chemical shift scales were calibrated using the unified $\Xi$ scale [17,18], relating the nuclear shift to the $^1$H resonance of dilute tetramethylsilane in CDCl$_3$ at a frequency of 300.13 MHz. The reference compounds for defining zero ppm on the chemical shift scales are tetramethyllead for $^{207}$Pb and dimethylselenide for $^{77}$Se.

### III. RESULTS AND DISCUSSION

Variable temperature $^{207}$Pb NMR spectra of PbSe acquired using the *v*ariable *o*ffset *c*umulative *s*pectra (VOCS) [19] technique are shown in Fig. 2. The ambient temperature spectrum shows a smaller resonance at -1460 ppm with a larger resonance at -2850 ppm. We shall henceforth use the terms "-2850 ppm" and "-1460 ppm" to refer to these two resonances.

Recently, a review article [20] on inorganic semiconductors states "if the range of isotropic chemical or Knight shifts is large, as it can be for the heavier nuclei such as [207]Pb [...] then MAS-NMR may offer no significant resolution increase". Previous [207]Pb NMR investigations of lead chalcogenides such as lead telluride (PbTe) [13] and other compounds [21] have shown that magic-angle spinning (MAS) fails to substantially narrow the NMR resonances. As shown in Fig. 3, a similar result for [207]Pb MAS is obtained for PbSe in this study. Due to the much larger breadth of the [207]Pb resonance in this sample of PbSe in comparison with that in Ref. 13, there is more loss of signal in refocusing the spin echo under MAS in this PbSe spectrum. There is an additional issue arising when using MAS to study semiconductor materials [20-22]. Specifically, the temperature of such conductive samples has been shown to be significantly higher than that expected only from frictional heating of the sample rotor as a result of spinning [13,20,22]. For these reasons, the NMR experiments were run on a static sample of PbSe. Aside from the difficulty of doing MAS in these samples, the presence of different resonances, as noted above, is confirmed by differences in the spin-lattice relaxation times (*vide infra*).

### A. NMR lineshapes

$^{207}$Pb spectra acquired in the range 173 K to 423 K are shown in Figs. 4 and 5. Since a single *RF* pulse cannot uniformly excite the entire $^{207}$Pb spectrum, appropriate changes of the *RF* offset frequency allow a "selective excitation" of each $^{207}$Pb resonance, except in the temperature region in which the resonances at -1460 ppm and -2850 ppm overlap, as shown in Fig. 6(a). In Fig. 4 the response of the larger resonance at -2850 ppm indicates a significant change in behavior below 250 K. The $^{207}$Pb chemical shift is linear with temperature above 250 K, as shown in Fig. 6(b), with a slope of 20.5 ppm/K. In addition, the linewidth narrows as the temperature is increased. An abrupt change in behavior below 250 K indicates the onset of an additional shift mechanism. To our knowledge, this change around 250 K is not discussed in the literature. The origins of these two mechanisms (above and below 250 K) are discussed below.

From the results of Fig. 6(b) it is clear that there is a different mechanism of the NMR shift of PbSe for the temperature dependence than that found in insulators, which exhibit a single mechanism [23]. The magnitude of the change in the shift as a function of temperature (20.5 ppm/K) is much larger than what has been typically observed in Pb-based diamagnetic materials such as 0.66 ppm/K for Pb(NO$_3$)$_2$ [20] or 1.61 ppm/K for PbI$_2$ [24]. Such a large temperature dependence has been observed in lead-based semiconductors, *e.g.*, the 10.1 ppm/K in PbTe [25]. These have been attributed to the nuclear spin-orbit term for *p*-type carriers and a dipolar field from the electrons (*n*-type), which can play a role even in cubic crystals [26] (e.g. PbTe, PbSe) owing to spin-orbit coupling.

Based on the classical expression of the NMR frequency shift in semiconductors, the temperature dependence of the NMR shift should obey the relation [26-32],

$$K = \frac{1}{3\hbar\pi^2}(2\pi)^{3/2} \cdot \gamma_e^2 \cdot \langle |u_{\vec{k}}(0)|^2 \rangle_{E_o} \cdot (m_e^* m_h^*)^{3/4} \cdot (k_B T)^{1/2} \cdot e^{-E_g/2k_B T}, \quad (1)$$

where $E_g$ is the energy gap, the $\langle |u_{\vec{k}}(0)|^2 \rangle_{E_o}$ denotes probability density of the single-particle carrier wavefunction at the nucleus, evaluated near the bottom of the conduction band for the electrons and the top of the valence band for holes [29-31], $\hbar$ is the Dirac constant, $\gamma_e$ the electron magnetogyric ratio and $m_e^*$, $m_h^*$ represent the effective masses of electrons and holes, respectively.

Lee *et al.* [14] suggested that the [207]Pb shift follows Eq. (1) across the entire temperature range. The linearity that they observed was explained as a single, uniform mechanism of the [207]Pb shift. However, such a simple temperature dependence of [207]Pb shift is not observed in our data. The frequency shift of -2850 ppm as function of inverse $T$ indicates instead a two-step process. Below 250 K, the shift becomes essentially independent of temperature, reaching a value of approximately -3000 ppm. The linewidth is also constant and about 3 times larger than that observed above 250 K [Fig. 6(b)]. The change in the NMR shift mechanism at 250 K observed in our experiments, which is not apparent in the data of Lee *et al.* [14], is likely a result of the better NMR resolution arising from the use of a higher magnetic field and a larger number of data points as a function of temperature than those collected by Lee and coworkers [14].

From analogy to the shifts in PbTe, the $^{207}$Pb shifts are expected to move downfield for *p*-type and upfield [33, 34, 14, 26] for *n*-type PbSe (Fig. 2-6).  However, in contrast to the -2850 ppm resonance tendency (Fig. 4), the resonance at -1460 ppm (Fig. 5) displays a smaller shift in the opposite direction as a function of temperature.  Based on the direction and magnitude of the temperature dependence of these two $^{207}$Pb shifts, the resonances should arise from *p*-type states for the case of the -2850 ppm and *n*-type states for the case of the -1460 ppm resonance.

The PbSe spectra show further significant changes at higher temperatures. Figure 6(c) shows an additional $^{207}$Pb resonance detectable only in the range 373 K to 423 K. At these temperatures, this new (third) $^{207}$Pb resonance appears at -147 ppm. We are unaware of any publications reporting the existence of this third resonance. As shown in Fig. 6(a), the reason for the limited temperature range in Fig. 5 is due to an overlapping of the -2850 ppm and -1460 ppm resonances as they move in different directions. On the other hand, the limited range in Fig. 6(c) arises for a different reason. It is not that the -147 ppm resonance begins to overlap with the -1460 ppm resonance.  The selective *RF* excitation of the -147 ppm peak shows that this peak actually "disappears" at temperatures below 373 K.

This raises a question as to whether the additional peak at -147 ppm might be a sign of irreversible (*e.g.*, annealing) effects on PbSe. Our previous study of PbTe [13] indicated sample annealing had occurred over this same temperature range as reflected by irreversible changes in both the $^{207}$Pb spectra and spin-lattice relaxation times following a thermal cycle. In contrast, PbSe in the current study did not show any signs of sample annealing over this same temperature range, a behavior that is consistent with the better thermal stability observed for PbSe compared to PbTe in power generation applications [7,35].

The observation of this third resonance above 370 K suggests a more complex semiconducting nature of PbSe than was hitherto reported. A resistivity study of PbSe thin films by Ali *et al.* [36] showed that the resistivity of PbSe displayed semiconductor behavior below 390 K but behaved as a metal above 400 K [36]. Although this new $^{207}$Pb resonance did not show a Korringa relationship characteristic of metallic behavior at temperatures above 370 K, this new resonance may be indicative of the change in resistivity behavior.

$^{77}$Se spectra of PbSe acquired from 173 K to 423 K are shown in Figure 7. In the plot of the $^{77}$Se chemical shift as a function of temperature in Fig. 6(d), the same change of behavior of the chemical shift below 250 K that was observed in Fig. 6(b) for $^{207}$Pb occurs. The results reveal a linear temperature dependence of $^{77}$Se with a clear change in slope at 250 K. The temperature dependence of the frequency shift is 0.3 ppm/K above 250 K and 0.15 ppm/K below 250 K. Similarly, the linewidth increases monotonically as the temperature decreases (-0.04 ppm/K). All $^{77}$Se spectra displayed symmetric lineshapes, which were well fitted with a Lorentzian function as shown in Fig. 7. The effect of conduction carriers on the $^{77}$Se nucleus (with *p*-like character) is much smaller than the case of the $^{207}$Pb nucleus (where carriers are *s*-like), therefore the presence of a

single component is expected [26]. Compared to $^{207}$Pb, the much lighter nucleus of $^{77}$Se hinders the visibility of the frequency change around the 250 K due to its weaker hyperfine coupling to the charge carriers.

## B. Spin-lattice relaxation rates

Spin-lattice relaxation ($T_1$) measurements for $^{207}$Pb versus temperature from 173 K to 423 K are shown in Fig. 8, 9 and for $^{77}$Se are shown in Fig. 10 (b). The saturation-recovery data of all resonances are fit well by a single exponential, $M(t) = M_0(1 - \exp(-t/T_1))$, where $M_o$ is the equilibrium magnetization and $t$ is the time elapsed after saturation. The single exponential decay is indicative of electronic homogeneity (see, *e.g.*, GeTe [12,37]). The electronic homogeneity as represented by a single component could be related to site-specific fractions of electrons or holes (*n*- or *p*-type semiconducting character) or a homogeneous distribution of native defects throughout the material (spread over a large region). The spin-lattice relaxation behavior of the three $^{207}$Pb resonances differs. For example, at 296 K the $^{207}$Pb spin-lattice relaxation time measured for the -2850 ppm resonance is 4.3 ms while that of the -1460 ppm resonance is 9.7 ms. In addition, from linear fitting of the $^{207}$Pb NMR (-2850 ppm) data we obtained $1/(T_1.T)$ = 2.37 s$^{-1}$K$^{-1}$ (T > 250 K) and $1/(T_1.T)$ = 1.19 s$^{-1}$K$^{-1}$ (T < 250 K) whereas the $^{207}$Pb resonances at -1460 ppm and -147 ppm gave values equal to 2.25 s$^{-1}$K$^{-1}$ and 3.64 s$^{-1}$K$^{-1}$, respectively. The decrease of $1/(T_1.T)$ with temperature is due to a reduction of the density of states (DOS) at the Fermi level as the temperature decreases for all the observed $^{207}$Pb resonances.

Dybowski et al. [38] developed a theory of nuclear spin-lattice relaxation mediated by magnetic coupling of the nuclear spins to lattice vibrations. This theory is applicable to several heavy spin-*1/2* nuclei including $^{207}$Pb, $^{119}$Sn and $^{203,205}$Tl. The Raman mechanism has been applied to $^{125}$Te and $^{77}$Se nuclei (both spin *I=1/2*) by Gunther and Kanert [31]. This Raman process [38, 39] involves the interactions between nuclear spins and lattice vibrations via a fluctuating spin-rotation magnetic field, which creates a relaxation pathway. The hallmark characteristics of this relaxation process are that the relaxation rate is proportional to $T^2$ and the relaxation rate extrapolates to zero at 0 K. Optical measurements on PbSe [35,40] have shown that optical phonons are gradually excited when approaching from above the Debye temperature ($T_\theta$ = 170 K). The acoustical phonons should appear close to 170 K.

In order to assess the role of the nuclear spin-rotation interaction [38], we plot the spin-lattice relaxation rates versus $T^2$. Figures 8(a) and 9(a) show the temperature dependence of the low temperature, *i.e.*, below 250 K, $^{207}$Pb spin-lattice relaxation rate as function of $T^2$ for the two main $^{207}$Pb resonances of PbSe. For comparison, our analysis follows the $^{207}$Pb-NMR study of Bouchard et al. [41] on PbTiO$_3$ and the $^{119}$Sn-NMR analysis by Neue et al. [42] on SnF$_2$, two well-studied compounds shown to follow the spin-rotation relaxation model. Our data extrapolated to low temperatures have an intercept of zero but do not agree with the $T^2$ prediction of the model. As shown in Figures 8 and 9, a Raman process mediated by spin-rotation interaction at these low temperatures is incompatible. Below 250 K, the larger resonance at -2850 ppm obeys a power-law mechanism, $1/(T_1.T) \propto T^a$ with exponent $a$ equal to 6.8 whereas above 250 K it is well fit by an exponential law, indicative of a thermally activated process.

The $^{207}$Pb spectra of the less intense resonance at -1460 ppm show a lower relaxation rate. Above 250 K, $T_1^{-1}$ rises rapidly following an exponential law over the entire temperature range, typical of a thermally activated mechanism observed in many semiconductors [Fig. 9(b)]. The third $^{207}$Pb resonance at -147 ppm reveals a similar behavior with the -1460 ppm resonance, in agreement with its spectral shift behavior. Furthermore, in the case of the -2850 ppm resonance, below 250 K the temperature dependence of the relaxation rate indicates a $T^{6.8}$ process, which may be related to the presence of acoustical phonon contributions ($T^7$) [40,43,44]. Long-wavelength acoustical phonons have been also implicated in optical and transport measurements of the same material [35,40]. This picture is at least partially consistent with our observation.

The $^{207}$Pb resonance at -1460 ppm, above 250 K shows a thermally activated mechanism with activation energy equals to 126.8 meV; nearly the half value of the direct band gap (0.27-0.31eV) of PbSe at the *L* point of the first Brillouin zone [1]. This activation energy (126.8 meV) of the *n*-type carriers may be related with the defect levels near the midgap [45] as currently proposed for PbTe by Ekuma *et al*. [40]. The extracted activation energy of 69.2 meV found for this resonance at -2850 ppm as shown in Fig. 10(a) matches the activation energy of 69.0 meV obtained from the $^{77}$Se $T_1$ data in Fig. 10(b), which is much smaller than the energy band gap of the PbSe (0.26-0.31 eV). This provides evidence for PbSe that the activation process of the *p*-type carriers (holes) may take place mainly through contributions from carriers that lie at the band edges or from localized electronic states in near to the valence band (shallow levels). The relaxation results indicate that PbSe relaxation process follows two mechanisms: a phonon-like contribution below 250 K and a thermally activated pathway above 250 K. As shown in Fig. 10, the $^{77}$Se and $^{207}$Pb dynamics are also described (T > 250 K) by an activated type of behavior.

## IV. CONCLUSIONS

Based upon our findings from $^{77}$Se and $^{207}$Pb NMR, PbSe emerges as a more complex semiconductor than previously reported. Namely, its electronic properties are spread among three distinct domains associated with three different $^{207}$Pb resonances. These resonances are due to discrete *n*- and *p*-carrier fractions within the PbSe matrix. A complex process, related to acoustical phonon contributions ($T^{6.8}$) should be considered accompanied with the involvement of carriers excited across the narrow bandgap. A noteworthy feature is that the activation energies of both nuclei ($^{207}$Pb and $^{77}$Se) above 250 K are found to be equal, revealing a common activation process for both sites that reflect the *p*-type states. In contrast, the $^{207}$Pb resonance peak at -1460 ppm which is related to *n*-type carriers revealed a thermal activation energy much higher (127 meV) than the aforementioned *p*-type states, almost equal to the half value of the entire bandgap (270 meV). A good understanding of the relaxation processes and hyperfine interactions is PbSe will be critical to the Pb$_{1-x}$Sn$_x$Se series, which describes a topological phase transition with *x* from topological crystalline insulator to normal insulator.

# ACKNOWLEDGMENTS

This research was supported by the Defense Advanced Research Project Agency (DARPA), Award No. N66001-12-1-4034.

**Figures and Captions**

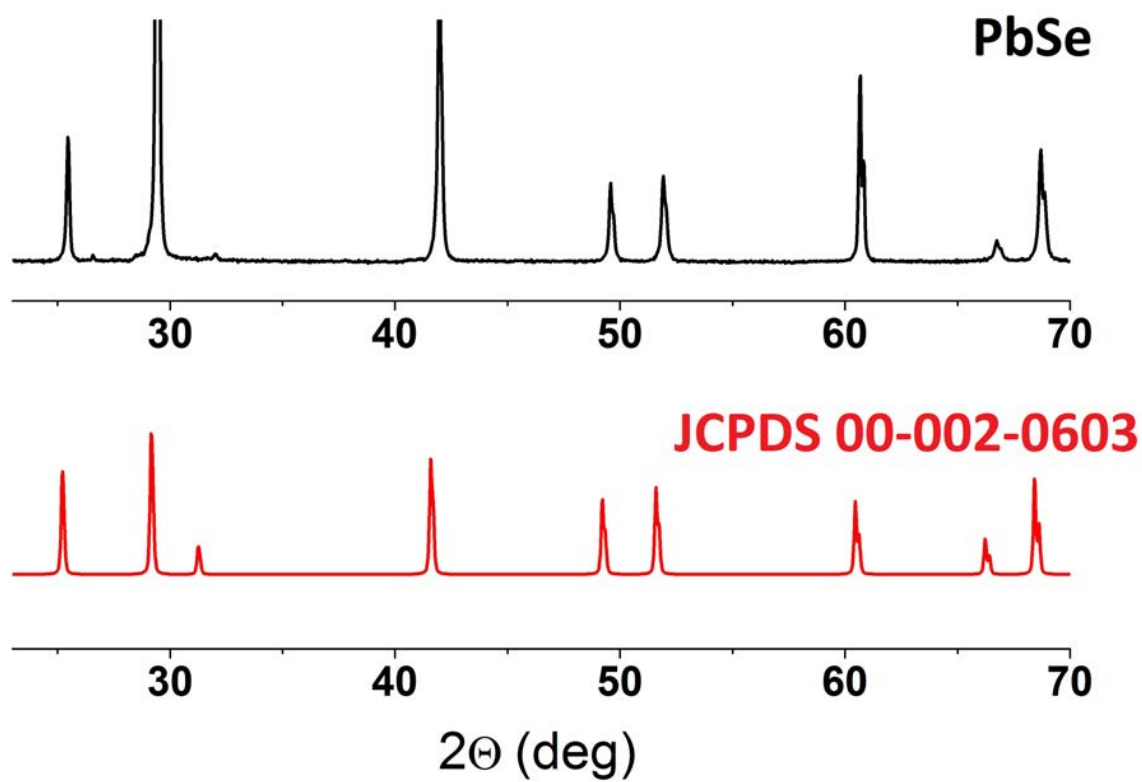

FIG. 1. (Color online) Powder X-ray diffraction spectra of lead selenide. The upper trace (black) is for PbSe (experimental, mortar & pestle). The red line is the PXRD spectrum of PbSe (JCPDS Reference). The PXRD spectra are given of the 20-80 deg region.

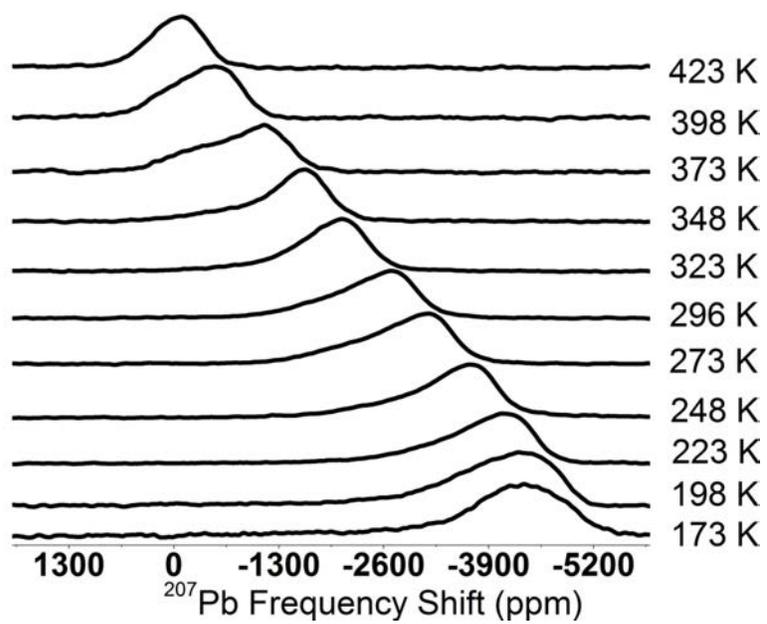

FIG. 2. $^{207}$Pb VOCS NMR spectra of PbSe over the temperature range of 173 K to 423 K.

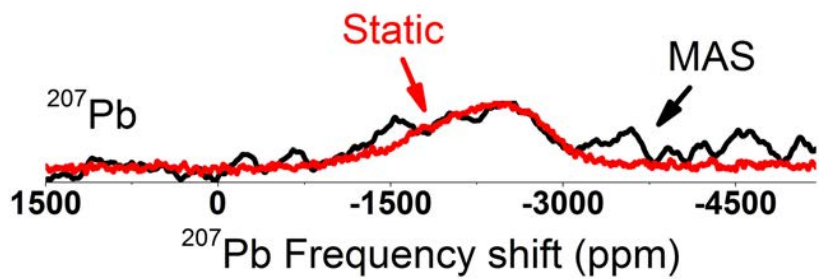

FIG. 3. (Color online) $^{207}$Pb spectrum (red line) of static polycrystalline PbSe at 295 K. The $^{207}$Pb MAS spectrum is as a black line. MAS performed with an 8 kHz spinning rate and does not narrow the static $^{207}$Pb resonance.

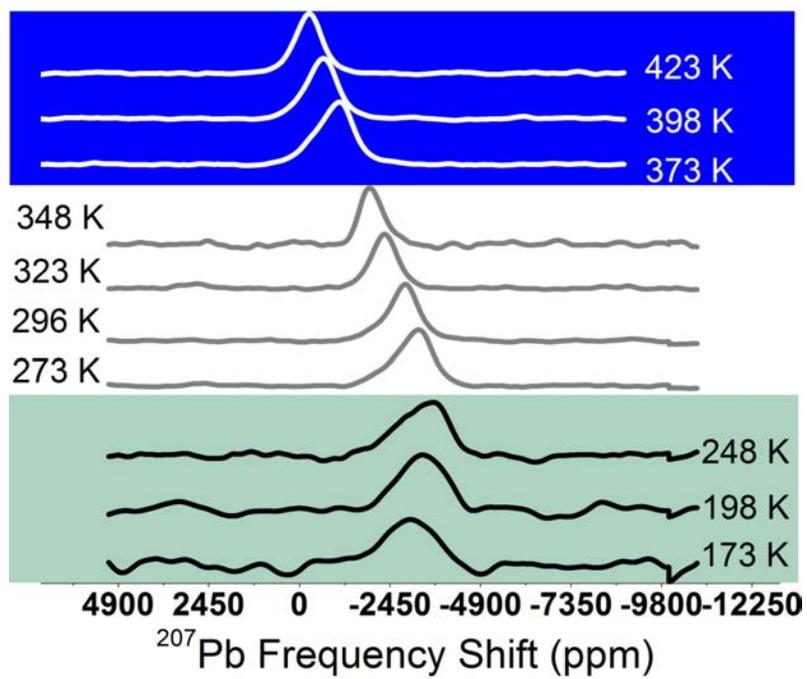

FIG. 4. (Color online) Selective *RF* excitation $^{207}$Pb (-2850 ppm) spectra of PbSe from 173 K to 423 K. Three different temperature regions are outlined.

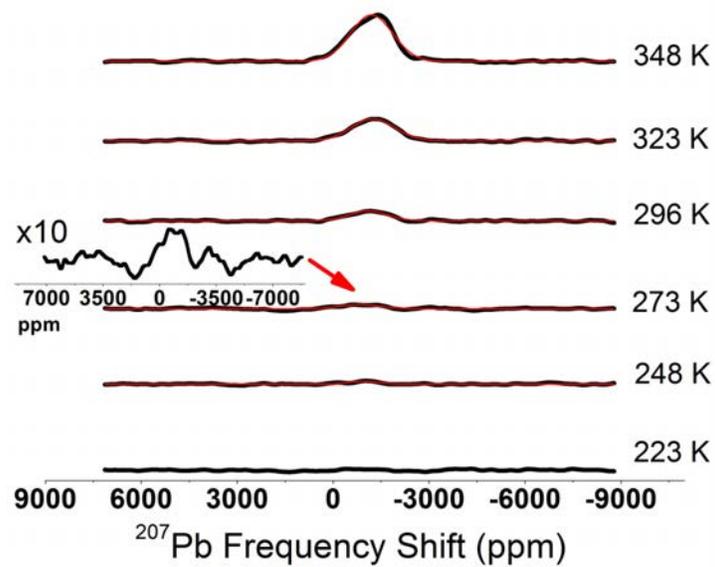

FIG. 5. (Color online) Selective *RF* excitation $^{207}$Pb (-1460 ppm) spectra of PbSe from 223 K to 348 K. The intensity drops as the linewidth broadens.

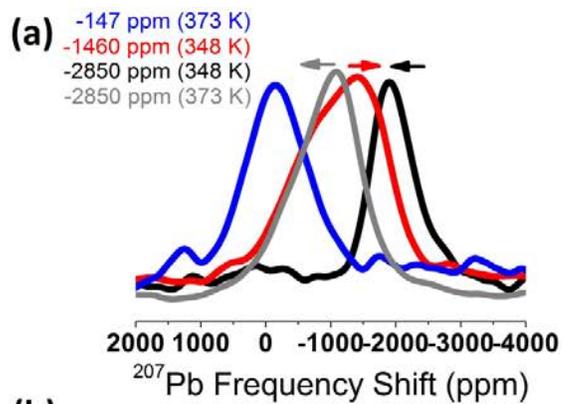
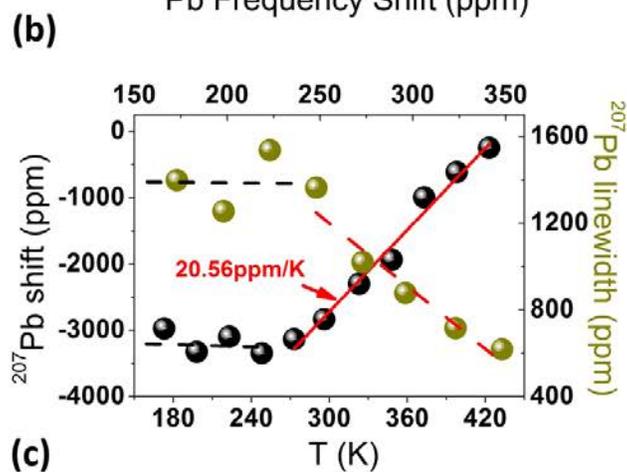
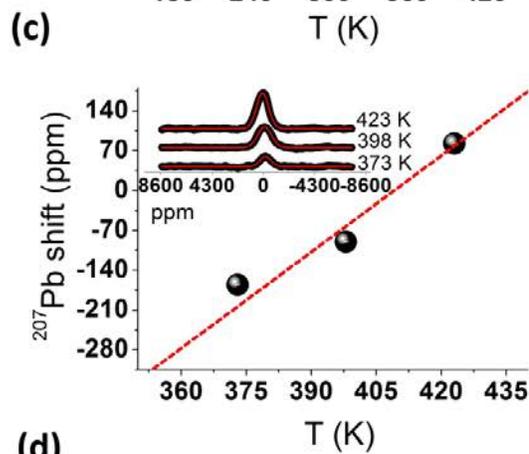
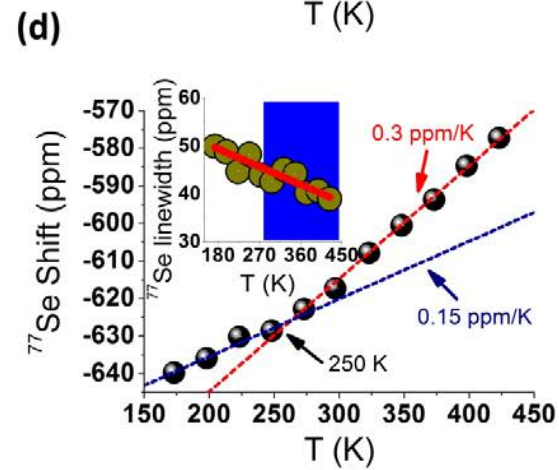

FIG. 6. (Color online) (a) NMR spectra collected at different *RF* offset frequencies reveal that in the high temperature regime (T>340 K), the resonance peaks at -2850 ppm (black curve, 348 K and grey curve, 373 K) and -1460 ppm (red curve, 348 K) begin to overlap as they move in different directions. A third resonance appears around -147 ppm (blue curve, 373 K). (b) $^{207}$Pb frequency shift analysis of the larger resonance at -2850 ppm of PbSe from 173 K to 423 K. The frequency shift as function of temperature shows a monotonic decrease down to 250 K. Below 250 K, the $^{207}$Pb resonance displays a constant value around -3000 ppm with a linewidth about 3 times larger. (c) Above 370 K, a third $^{207}$Pb resonance appears around -147 ppm. $^{207}$Pb (-147 ppm) spectra of PbSe from 373 K to 423 K are shown whereas the upper plot shows its temperature-dependent resonance shift. (d) $^{77}$Se frequency shift analysis of PbSe from 173 K to 423 K. The data reveal a strong linear temperature dependence of $^{77}$Se with a cusp at 250K. Above the 250 K the monotonic decrease of the frequency shift gives a value close to 0.3 ppm/K whereas below 250 K is equal to 0.15 ppm/K. Additionally, the linewidth increases monotonically as the temperature decreases (-0.04 ppm/K).

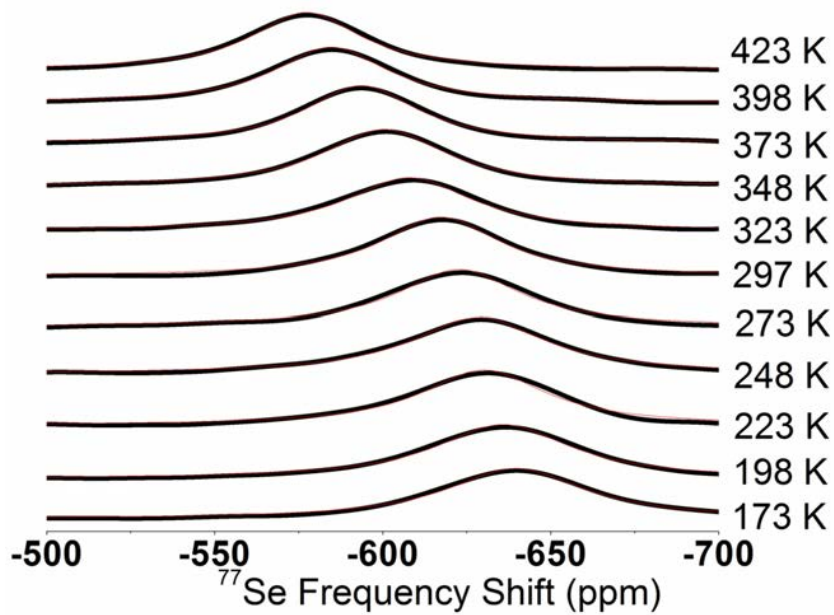

FIG. 7. $^{77}$Se spectra of PbSe from 173 K (bottom) to 423 K (top). $^{77}$Se spectra remained symmetric over the entire temperature range. The $^{77}$Se resonance follows a linear temperature dependence with a slope change at 250 K.

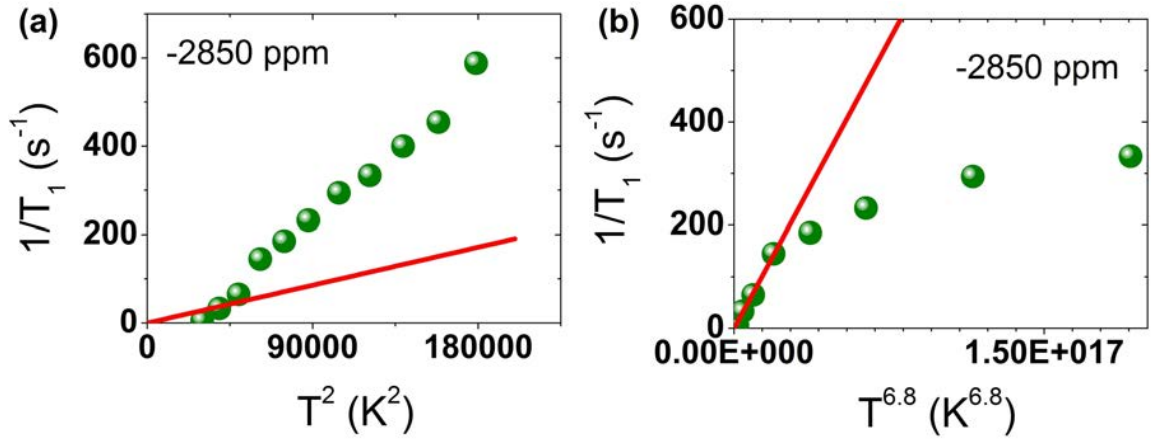

FIG. 8. (Color online) Spin-lattice relaxation rate vs. $T^2$ for the -2850 ppm $^{207}$Pb resonance of PbSe (a). The red straight line reflects the expected behavior due to a Raman process mediated by spin-rotation interaction. Below 250 K, the $1/T_1$ of the $^{207}$Pb (-2850 ppm) resonance is well described by a $T^{6.8}$ law, as shown by the red straight line (b).

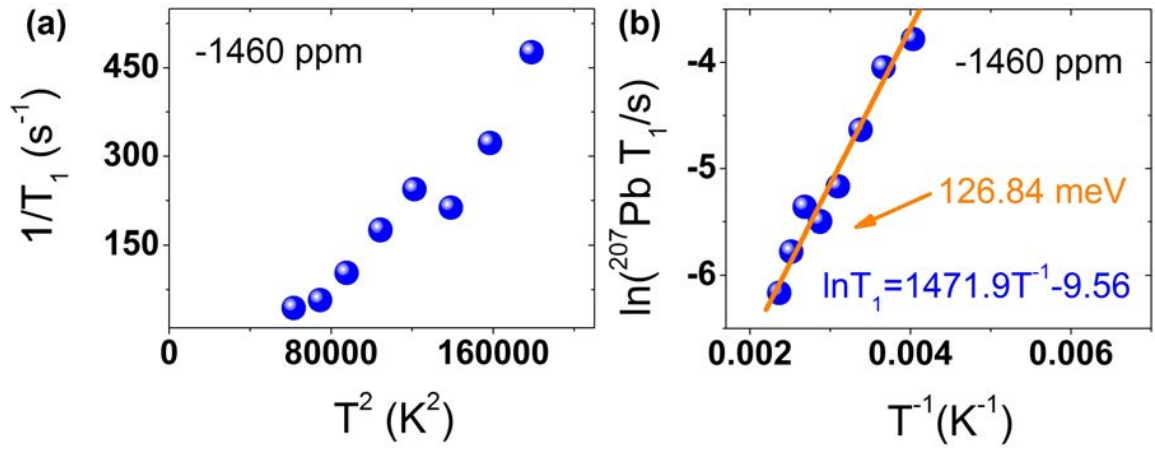

FIG. 9. (Color online) Spin-lattice relaxation rate vs. $T^2$ for the -1460 ppm $^{207}$Pb resonance of PbSe (a). The $^{207}$Pb (-1460 ppm) resonance at the high temperature region is described by a thermally activated mechanism (126.8 meV) (b).

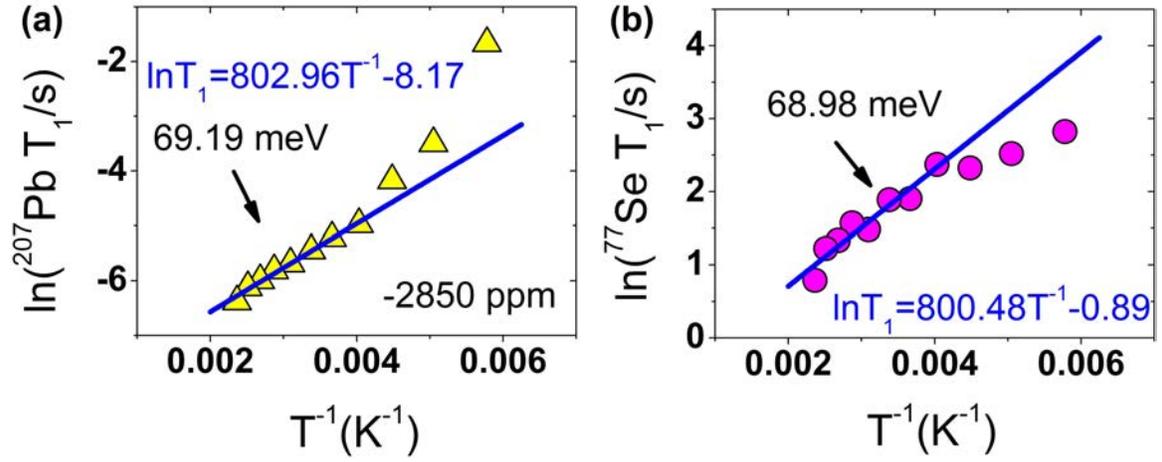

FIG. 10. (Color online) (a) Semilogarithmic plot of the $^{207}$Pb (-2850 ppm) and $^{77}$Se spin-lattice relaxation rate for PbSe from 173 K to 423 K (b). The activation energy of 69.2 meV found for $^{207}$Pb (-2850 ppm) resonance matches the activation energy of 69.0 meV obtained from the $^{77}$Se analysis.

*Corresponding author (L.-S. B) address: Department of Chemistry and Biochemistry, University of California, Los Angeles, USA. Electronic address: bouchard@chem.ucla.edu